\documentclass[twocolumn,aps,prb]{revtex4}
\usepackage[latin9]{inputenc}
\usepackage{amsbsy}
\usepackage{amstext}

\makeatletter

\usepackage{bm}
\usepackage{graphicx}
\usepackage{epsfig}

\makeatother

\begin{document}
\title{Anisotropic superexchange through nonmagnetic anions with spin-orbit
coupling}
\author{Jun Chang$^{1}$, Jize Zhao$^{2}$, Yang Ding$^{3}$}
\affiliation{$^{1}$College of Physics and Information Technology, Shaanxi Normal
University, Xi'an 710119, China~\\
 $^{2}$School of Physical Science and Technology\&Key Laboratory
for Magnetism and Magnetic Materials of the MoE, Lanzhou University,
Lanzhou 730000, China~\\
 $^{3}$Center for High-Pressure Science and Technology Advanced Research,
Beijing 100094, China~}
\date{\today}
\begin{abstract}
Anisotropic superexchange interaction is one of the most important interactions in realizing exotic quantum magnetism, 
which is traditionally regarded to originate from magnetic ions and has no relation with
the nonmagnetic ions. In our work, by studying a multi-orbital Hubbard
model with spin-orbit coupling on both magnetic cations and nonmagnetic
anions, we analytically demonstrate that the spin-orbit coupling on
nonmagnetic anions alone can induce antisymmetric Dzyaloshinskii-Moriya
interaction, symmetric anisotropic exchange and single ion anisotropy
on the magnetic ions and thus it actually contributes to anisotropic
superexchange on an equal footing as that of magnetic ions. Our results
promise one more route to realize versatile exotic phases in condensed
matter systems, long-range orders in low dimensional materials and
switchable single molecule magnetic devices for recording and manipulating
quantum information through nonmagnetic anions. 
\end{abstract}
\maketitle

\section{\label{intro}Introduction}

Locking electron spin and momentum together, the relativistic spin-orbit
coupling (SOC) plays a critical role in realizing a diversity of exotic
phases in condensed matter systems, such as quantum spin liquid, spin-orbit
coupled Mott insulator, Weyl semimetal and topological insulator \cite{Witczak-Krempa2014,Bansil2016,RevModPhys.87.1}.
In the absence of SOC, the magnetic interaction is isotropic with
spin rotational invariance. However, SOC may lower the symmetry and
lead to anisotropic interactions, which has been microscopically
identified by Moriya by means of extending the Kramers-Anderson superexchange
theory \cite{Moriya1960,Anderson1959,Kanamori1963,Goodenough1968}.
Importantly, the magnetic anisotropy is the key in bond-dependent
Kitaev interaction and phase transition in low dimensional (D$\leq$2)
systems, which has been argued to be a great promise for quantum computation
and information processing in addition to the fundamental interest \cite{Kitaev2006,Chen2008,Jackeli2009}.
\begin{figure}
\includegraphics[width=0.9\columnwidth]{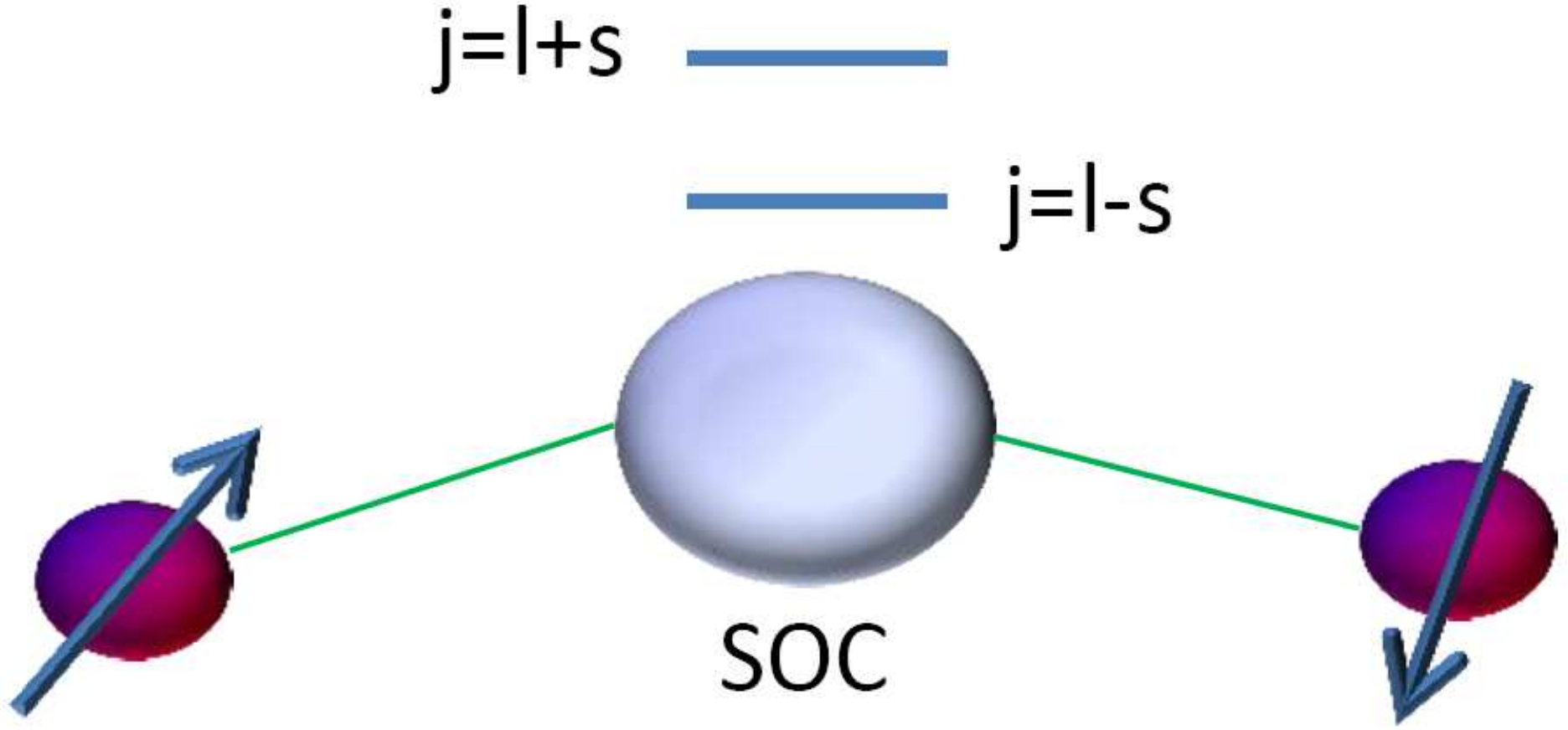} \caption{Superexchange between two magnetic ions via a nonmagnetic ligand with spin-orbit interaction. Apart from the crystalline electric field splitting, the SOC on the nonmagnetic ligands could further split the energy levels in two fine structures with the spin moment parallel and anti-parallel to the orbital moment, respectively. The SOC on ligands could induce spin flip in the virtual hopping process of superexchange. 
}
\label{superexchange} 
\end{figure}

Recently, magnetic orders induced by magnetic aniso-tropy have been
reported in experiments on some two-dimensional (2D) materials or
proposed by numerical simulations 
\cite{Chen2018,Gong2017,Seyler2018,Xu2018,Lado2017,Huang2017}.
Since it is well known that the Hohenberg-Mermin-Wagner theorem forbids
the spontaneous breaking of continuous symmetry at finite temperature
in low dimensional systems \cite{Mermin1966}, these findings are thus
attributed to magnetic anisotropy in the materials. Commonly, the
magnetic anisotropy is believed to dominantly result from the SOC
on magnetic ions. Nevertheless, intuitively, the SOC on the nonmagnetic
anions should also induce spin flip in the virtual hopping process
of superexchange. Indeed, in ferromagnetic CrI$_{3}$ and CrGeTe$_{3}$
monolayers, an-isotropic exchange coupling and single ion anistropy
are found to be dominated by the SOC from the nonmagnetic 5$p$ Iodine
or Tellurium anions rather than the magnetic 3$d$ Cr ions \cite{Lado2017,Xu2018}.
Although evidences have been provided numerically by comparing the
relevant magnetic couplings with and without SOC \cite{Lado2017,Xu2018}, from the theoretical point
of view, an insight into the microscopic mechanism to understand how the SOC on nonmagnetic ions induces the magnetic anisotropic superexchange on magnetic ions is of both importance and interest \cite{Winter2017,Jackeli2009,RevModPhys.87.1}. Actually, some relavent studies have been started since 1980s, which focuse on the RKKY interaction via the conduction electron with spin-orbit coupling \cite{Fert1980,Imamura2004}.

In this paper, we consider a general system in which magnetic ions
interact through the superexchange via nonmagnetic ions with SOC on
the latter. Starting from a multi-orbital model with orbital hybridization,
onsite Cou-lomb interaction and SOC on both magnetic cations and nonmagetic
anions, we first transform this model into a multiorbital model with
both spin-isotropic and spin-anisotropic electron hopping integrals
between magnetic and nonmagnetic ions. Next, applying the degenerate
perturbation theory, we obtain the isotropic exchange, aniso-tropic
exchange or pseudodipolar interaction, antisymmetric Dzya-loshinskii-Moriya (DM)
interaction and single-ion anisotropy on magnetic ions. We show that
the SOC on nonmagnetic ions contributes to magnetic anisotropy in
a similar way as that on the magnetic ions. Moreover, the anisotropic
magnetic couplings increase sharply with the atomic number. Therefore,
the heavy nonmagnetic ligands may contribute dominantly to the magnetic
anisotropy rather than the light magnetic ions.

\section{ Microscopic Hamiltonian}

We model a general system containing magnetic cations M surrounded by
nonmagnetic anions or ligands NM. The central interactions include
the spin-orbit angular momentum coupling $\mathbf{L}\cdot\mathbf{s}$
on both M and NM ions, and the onsite Coulomb interaction $U_{d}$ and $U_{p}$ between electrons on the M and NM ions, respectively.
The perturbation term is assumed to be the electron transfer integral
$t_{imkn}$ between the $m$th orbital of M ion at site $i$ and the
$n$th orbitals of the ligand NM at site $k$. We neglect the direct
hopping between M sites. Then, the Hamiltonian can be written as 
\begin{eqnarray}
H=H_{d}+H_{p}+H_{pd},\label{H}
\end{eqnarray}
where 
\begin{eqnarray}
H_{c=p,d}=H_{c0}+\frac{1}{2}\sum_{imn\alpha\beta}U_{cm\alpha n\beta}n_{cim\alpha}n_{cin\beta},
\end{eqnarray}
\begin{eqnarray}
H_{pd}=\sum_{ikmn\alpha}\left(t_{imkn}d_{im\alpha}^{\dagger}p_{kn\alpha}+{H.c.}\right),\label{Hpd}
\end{eqnarray}
with 
\begin{eqnarray}
H_{c0}=\sum_{im\alpha n\beta}E_{m\alpha,n\beta}^{ci}c_{im\alpha}^{\dagger}c_{in\beta},
\end{eqnarray}
\begin{eqnarray}
E_{m\alpha,n\beta}^{ci}=\varepsilon_{cm0}\delta_{mn}\delta_{\alpha\beta}+{\ensuremath{\lambda}}_{c}\left\langle l_{c}m\alpha\left|\mathbf{L}_{ci}\cdot\mathbf{s}\right|l_{c}n\beta\right\rangle ,
\end{eqnarray}
where $d_{im\alpha}^{\dagger}$ is the creation operator of an electron
with spin $\alpha$ on the $m$th orbital of $i$th M ion, and $\varepsilon_{dm0}$
is the energy of the $m$th orbital. We assume that the crystalline
electric field effect has been taken into account by lifting the degeneracy
of the orbital energy levels. $p_{kn\alpha}^{\dagger}$ is the creation
operator of an electron on the $m$th orbital with spin $\alpha$.
$n_{d}$ and $n_{p}$ are the particle number operators of the electrons
on M and NM sites, respectively. $\mathbf{L}$ and $\mathbf{s}$ denote
the orbital and spin angular momenta. $\lambda_{d}$ and $\lambda_{p}$
are the spin-orbit coupling constants on the M and NM ions.

It is easy to diagonalize the single-site Hamiltonian matrix $E^{ci}$
firstly \cite{Chen2008,Yildirim1995}. In general, we have a site-dependent
diagonal eigenvalue matrix $\varepsilon_{ic}=W^{ci\dagger}E^{ci}W^{ci}$ with
the unitary eigenvector matrix $W^{ci}$. Choosing real orbital wave
functions, then the $\mathbf{L}_{c}$ matrix elements are purely imaginary.
$\varepsilon_{ci}$ has $l_{c}$ eigenvalues with the orbital angular
momentum quantum number $l_{c}$ and every single-particle energy
of $\varepsilon_{ci}$ has at least a twofold degeneracy due to time
reversal symmetry, characterized by pseudospin quantum numbers, $\pm1$/2.
Consequently, a new set of electron creation and annihilation operators
is introduced for $p$ and $d$ electron operators 
\begin{eqnarray}
q_{kn\alpha}=\sum_{l\beta}W_{n\alpha,l\beta}^{pk}p_{kl\beta}, a_{im\alpha}=\sum_{n\beta}W_{m\alpha,n\beta}^{di}d_{in\beta}.
\end{eqnarray}
Let's express the microscopic Hamiltonian in terms of $q_{km\alpha}$
and $a_{im\alpha}$ as 
\begin{eqnarray}
H_{p}=&&\sum_{kn\alpha}\varepsilon_{pkn}n_{qkn\alpha}\nonumber \\ &&+\frac{1}{2}\sum_{if\rho h\sigma l\tau r\upsilon}V_{f\rho h\sigma l\tau r\upsilon}^{pk}q_{kf\rho}^{\dagger}q_{kh\sigma}q_{kl\tau}^{\dagger}q_{kr\upsilon},
\end{eqnarray}

\begin{eqnarray}
H_{d}=&&\sum_{im\alpha}\varepsilon_{dim}n_{aim\alpha}\nonumber \\ &&+\frac{1}{2}\sum_{if\rho h\sigma l\tau r\upsilon}V_{f\rho h\sigma l\tau r\upsilon}^{di}a_{if\rho}^{\dagger}a_{ih\sigma}a_{il\tau}^{\dagger}a_{ir\upsilon},
\end{eqnarray}

\begin{eqnarray}
H_{pd}=\sum_{imkn\alpha}a_{im\alpha}^{\dagger}(b_{imkn}\delta_{\alpha\beta}+\mathbf{C}_{imkn}\cdot\boldsymbol{\sigma}_{\alpha\beta})q_{kn\beta}\nonumber \\
+{H.c.}
\end{eqnarray}
with 
\[
V_{f\rho h\sigma l\tau r\upsilon}^{pk}=\sum_{mn\alpha\beta}U_{pm\alpha n\beta}W_{m\alpha,f\rho}^{pk}W_{m\alpha,h\sigma}^{pk\dagger}W_{n\beta,l\tau}^{pk}W_{n\beta,r\upsilon}^{pk\dagger},
\]
\[
V_{f\rho h\sigma l\tau r\upsilon}^{di}=\sum_{mn\alpha\beta}U_{dm\alpha n\beta}W_{m\alpha,f\rho}^{di}W_{m\alpha,h\sigma}^{di\dagger}W_{n\beta,l\tau}^{di}W_{n\beta,r\upsilon}^{di\dagger},
\]
 the spin-isotropic hopping matrix elements 
\begin{eqnarray}
b_{imkn}=\frac{1}{2}\sum_{\alpha\gamma m'n'}W_{m\alpha,m'\gamma}^{di}t_{im'kn'}W_{n'\gamma,n\alpha}^{pk\dagger},
\end{eqnarray}
and the spin-anisotropic matrix elements 
\begin{eqnarray}
\mathbf{C}_{imkn}=\frac{1}{2}\sum_{\alpha\beta\gamma m'n'}W_{m\alpha,m'\gamma}^{di}t_{im'kn'}W_{n'\gamma,n\beta}^{pk\dagger}\mathbf{\boldsymbol{\sigma}_{\alpha\beta}}.
\end{eqnarray}
where $\mathbf{\boldsymbol{\sigma}}$ is the vector of the three Pauli
matrix. Clearly, if we ignore the SOC on nonmagnetic anions, namely,
setting $W^{pk}$ a unit matrix, Moriya's result is reproduced \cite{Moriya1960,Yildirim1995,Chen2008}.
On the other hand, ignoring the SOC on the magnetic cations, i.e. setting
$W^{di}$ a unit matrix, we still reach a similar result. Formally,
the SOC on nonmagnetic ions contributes to both the spin-isotropic
and spin-anisotropic hopping matrix elements in a similar way to that
on the magnetic ions. In the weak SOC limit, the spin-anisotropic
hopping matrix elements are in the linear order of $\lambda_{c}/\Delta\varepsilon_{c}|_{c=p,d}$
with the energy difference $\Delta\varepsilon_{c}$ between the ground
state and the excited state.

\section{Superexchange Interaction}

For magnetic materials, it is often assumed that localized electrons
yield both orbital and spin magnetic moments on each ions or part
of ions. The direction of a magnetic moment is described by a classical
unit vector. It is natural to define an effective spin to match up
with the local magnetization. Here, we define effective spin operators
in terms of the creation $a_{im}^{\dagger}$ and annihilation $a_{im}$
operators as 
\begin{eqnarray}
\mathbf{S}_{i}=\frac{1}{2}\sum_{\alpha\beta m}a_{im\alpha}^{\dagger}\boldsymbol{\sigma}_{\alpha\beta}a_{im\beta}.
\end{eqnarray}
Due to the time-reversal symmetry, the fully occupied pseudospin levels
give no contribution to magnetization. Therefore, the magnetization
is dominated by the half-occupied effective spin levels. Moreover,
on magnetic atoms, the electron kinetic energy or $H_{pd}$ is often weak enough
to be regarded as perturbation, and then the generalized Hubbard Hamiltonian
$H$ in Eq. (\ref{H}) is reduced to an effective spin interaction
Hamiltonian of magnetic ions, mediated by the nonmagnetic ions. To obtain the spin interaction between magnetic
ions M, it is customary to appeal to the degenerate perturbation theory
by taking the hybridization $H_{pd}$ or hopping as the perturbation.
We consider the case that the hopping is much less than onsite Coulomb
interactions $U_{p,d}$ and the charge transfer energy between NM and
M ions. A common way is to use the limit $U_{d}\rightarrow\infty$
to forbid double occupancy in the magnetic atom orbitals. According
to Goodenough-Kanamori rule, half-occupied orbitals dominate the exchange
coupling and fully occupied orbitals' contribution is much weaker \cite{Goodenough1968,Kanamori1963}.
We define a projection operator $\hat{P}_{g}$, which projects onto
the subspace with only single electron locating on each pseudospin
levels of magnetic ions, and another projector $\hat{P}_{e}=\hat{I}-\hat{P}_{g}$
with the unit operator $\hat{I}$. We take the hopping term $H_{pd}$
as the perturbation to derive an effective Hamiltonian \cite{Yildirim1995,Chang2017}.
The $n$th order perturbation reads, 
\begin{eqnarray}
H^{(n)}=\hat{P}_{g}H_{pd}(\frac{1}{E-(H_{d}+H_{p})}\hat{P}_{e}H_{pd})^{n-1}\hat{P}_{g},\label{Hn}
\end{eqnarray}
with the ground state energy $E$. One finds the first and the third
order perturbations vanish because odd-order hops introduce double
or empty occupancy on the originally half-filled magnetic atom orbitals, i.e. $H^{(1)}=H^{(3)}=0$.
The second-order contribution is dominated by the magnetic coupling
between electrons on NM and M sites, e.g. $\mathbf{S}_{pk}\cdot\mathbf{S}_{di}$.
However, this term gives no contribution because the total magnetic
moment on each NM ion is zero in the ground state, or $\left\langle \mathbf{S}_{pk}\right\rangle =0$.
Consequently, the leading contribution to superexchange is the fourth
order in hopping between M and NM ions, which consists of the ``hops''
described by spin-isotropic matrix elements $b_{ik}$, and the ``hops''
given by anisotropic $C_{ik}$ matrix elements. We have assumed that
the electron number fluctuation on the NM ions is weak enough so that
it can be ignored.

In time-reversal invariant systems, the magnetic Hamiltonian consists
of two-spin interactions between nearest and further neighbors, four
spin interactions, and so forth \cite{Yildirim1995,Anderson1959,Moriya1960,Takahashi1977}.
In this paper, we confine to the two-spin interactions. A general
bilinear two-spin interaction can be written as 
\begin{eqnarray}
H_{ex}=\sum_{ij}\mathbf{S}_{i}\cdot\mathbf{M}_{ij}\cdot\mathbf{S}_{j}.\label{Hex1}
\end{eqnarray}
Usually, the $3\times3$ interaction-matrix $\mathbf{M}_{ij}$ is
separated into an antisymmetric and a symmetric matrix 
\begin{eqnarray}
\mathbf{M}_{ij}^{\mp}=\frac{\mathbf{M}_{ij}\mp\mathbf{M}_{ij}^{T}}{2}.
\end{eqnarray}
Further, the symmetric $\mathbf{M}_{ij}^{+}$ is often split into
an isotropic coupling matrix $J_{ij}\mathbf{I}$, and a symmetric
traceless one 
\begin{eqnarray}
\mathbf{\Gamma}_{ij}=\mathbf{M}_{ij}^{+}-J_{ij}\mathbf{I}
\end{eqnarray}
where $J_{ij}=Tr(\mathbf{M}_{ij})/3$ and $\mathbf{I}$ is a unit
matrix. Alternatively, the eigenvalues of the symmetric matrix $\mathbf{M}_{ij}^{+}$
are obtained by diagonalization, and then separated into the isotropic
exchange and Kitaev coupling \cite{Xu2018}. The antisymmetric exchange
is commonly written as DM interaction $\mathbf{D}_{ij}\cdot\left(\mathbf{S}_{i}\times\mathbf{S}_{j}\right)$
since each antisymmetric $3\times3$ matrix can be linearly mapped
onto a 3D vector, $(\mathbf{M}_{ij}^{-})_{\mu\nu}=\sum D_{ij,\lambda}\varepsilon_{\lambda\mu\nu}$
with Levi-Civita symbol $\varepsilon_{\lambda\mu\nu}$ and Cartesian
components $\lambda$, $\mu$ and $\nu$. Thus, the exchange Hamiltonian
is rewritten in terms of the isotropic Heisenberg exchange, DM interaction,
and symmetric anisotropic exchange as 
\begin{eqnarray}
H_{ex}=\sum_{ij}J_{ij}\mathbf{S}_{i}\cdot\mathbf{S}_{j}+\mathbf{D}_{ij}\cdot\left(\mathbf{S}_{i}\times\mathbf{S}_{j}\right)+\mathbf{S}_{i}\cdot\mathbf{\Gamma}_{ij}\cdot\mathbf{S}_{j}.\nonumber \\
\label{Hex2}
\end{eqnarray}

Casting off some constants and hopping terms, the $4$th order perturbation
$H^{(4)}$ in equation (\ref{Hn}) is rearranged into $H_{ex}$ form.
The interaction matrix tensors in $H_{ex}$ include the isotropic
Heisenberg coupling 
\begin{eqnarray}
J_{ij}=4\sum_{knk'n'}s_{ijkn}g_{knk'n'}s_{jik'n'},
\end{eqnarray}
Dzyaloshinskii-Moriya vectors 
\begin{eqnarray}
\mathbf{D}{}_{ij}=-4i\sum_{knk'n'}g_{knk'n'}\left(\mathbf{v}_{ijkn}s_{jik'n'}-s_{ijkn}\mathbf{v}_{jik'n'}\right),\nonumber \\
\end{eqnarray}
and symmetric anisotropic exchange 
\begin{eqnarray}
\mathbf{\mathbf{\Gamma}}_{ij}= & 4 & \sum_{knk'n'}g_{knk'n'}\left(\mathbf{v}_{ijkn}\mathbf{v}_{jik'n'}+\mathbf{v}_{jikn}\mathbf{v}_{ijk'n'}\right)\nonumber \\
 & - & g_{knk'n'}\mathbf{I}\left(\mathbf{v}_{ijkn}\cdot\mathbf{v}_{jik'n'}\right)
\end{eqnarray}
with unit matrix $\mathbf{I}$. To simplify the expressions of the
coupling matrix, the effective spin-isotropic hopping matrix elements
have been introduced between $i$ and $j$ magnetic ions after the
definition by Chen and Balents \cite{Chen2008} 
\begin{eqnarray}
s_{ijkn}=b_{ikn}b_{knj}+\mathbf{C}_{ikn}\cdot\mathbf{C}_{knj},
\end{eqnarray}
the effective spin-anisotropic hopping matrix elements 
\begin{eqnarray}
\mathbf{v}_{ijkn}=\mathbf{C}_{ikn}b_{knj}+b_{ikn}\mathbf{C}_{knj}+i\left(\mathbf{C}_{ikn}\times\mathbf{C}_{knj}\right),
\end{eqnarray}
and the coefficients originating from the third power of $(E-H_{d}-H_{p})^{-1}$
in the $4$th order perturbation $H^{(4)}$ 
\begin{eqnarray}
g_{knk'n'}=&&\frac{\left(1-\frac{1}{2}\delta_{kk'}\delta_{nn'}\right)}{E_{knk'n'}}\left(E_{kn}^{-1}+E_{k'n'}^{-1}\right)^{2}\nonumber \\
 && + \frac{1}{E_{kn}E_{j}E_{k'n'}},
\end{eqnarray}
where $E_{kn}$ is the energy difference between the ground state
and the intermediate excited state with one electron or hole hopping
to the nth orbital on $k$ site from the $i$ or $j$ site. Commonly,
as a function of $\varepsilon_{d}$ and $\varepsilon_{p}$ as well
as the onsite Coulomb interaction $U_{d}$ and $U_{p}$, the detail
value of $E_{kn}$ depends on the electron configuration. Similarly,
$E_{j}$ is the single particle excitation energy for the $j$ ions.
$E_{knk'n'}$ corresponds to the double-electron or hole activation
energy for two electrons or holes hopping to the nth orbital on the
$k$ site and the $n'$th orbital on the $k'$ site from $i$ and
$j$ site.

In addition, the SOC on nonmagnetic ligands could also induce the
local spins on magnetic ions to align along a specific local axis,
i.e. the single-ion anisotropy or magnetocrystalline anisotropy. The
single-ion anisotropy could be of the same order in magnitude as the
pseudodipole or Kitaev interactions \cite{Xu2018} . It is written
as a lattice-dependent onsite tensor 
\begin{eqnarray}
\mathbf{\mathbf{\Gamma}}_{ii}=4\sum_{knk'n'}2\mathbf{v}_{iikn}g_{knk'n'}\mathbf{v}_{iik'n'}\nonumber \\
-g_{knk'n'}\mathbf{I}\left(\mathbf{v}_{iikn}\cdot\mathbf{v}_{iik'n'}\right).
\end{eqnarray}
This provides a novel route to engineer the single-molecule magnetic
switch devices for storing or manipulating information with the aid
of spins. 

\section{Discussion and Conclusion}

DMI and anisotropic exchange as well as the single ion anisotropy depend on the geometrical symmetry of lattice. The introduction of the SOC effect of ligands actually contributes to anisotropy on an equal footing as that of magnetic ions via the hopping matrix, and maintains the symmetries of both the hopping matrix elements and crystal field levels. Therefore, the dependence of the magnetic anisotropy on the crystal symmetry is just the same as that of the SOC effects on the magnetic ions. The DMI and anisotropic exchange are still anti-symmetric and symmetric, respectively, e.g. $D_{ij} = - D_{ji}$ and $\Gamma_{ij} =\Gamma_{ji}$. For example, in a system without inversion, both DMI and anisotropic exchange could emerge, depending on the detail geometric lattice symmetry. On the other side, in a system with inversion symmetry, DMI disappears but the anisotropic exchange could be preserved. If the symmetry of the hopping matrix elements alters, then the magnetic anisotropy may change accordingly. For instance, a magnetic ion has no single ion anisotropy in a cubic lattice surrounded by 6 same ligands. When one ligand is replaced by a nonmagnetic ion with higher SOC and even the crystalline electric field is kept the same, the single ion anisotropy is induced with the magnetic moment vector alone the line connecting the magnetic ion and the substituted ligand. The original geometric symmetry is broken due to the different SOC of the substituted ligand and hence the hopping matrix elements. 

Although we assume that the single-site energies $\varepsilon_{ic}$
may depend on the site index, in certain situations, for example,
with the same crystalline electric field for each magnetic ion, it
is possible to define site-dependent transformations $W^{ic}$ such
that the single-particle energies are site independent. However, the
overlaps between orbital states, e.g. the spin-anisotropic matrix
elements $\mathbf{C}_{imjn}$ and the symmetric anisotropic exchange
$\mathbf{\mathbf{\Gamma}}_{ij}$ could be bond-dependent due to the
orbital anisotropy and crystal lattice geometry, including frustration \cite{Wang2011}.
Therefore, taking advantage of the properties of the heavy nonmagnetic
ligands, additional methods are provided to design potential exotic
spin models, such as the Kitaev model with bond-dependent anisotropic
interaction.

Within Moriya's theoretical framework, the DM interaction and the
anisotropic exchange are proportional to $\lambda_{c}$ and their
quadratic forms, respectively. Moreover, $\lambda_{c}$ increases
sharply with the atomic number $Z_{c}$, namely, $\lambda_{c}\propto Z_{c}^{4}$.
Since the SOC on both the nonmagnetic ligands and the magnetic ions
contributes in the same way to the magnetic anisotropy, the magnetic
anisotropy is thus dominated by the heavier ions. For example, in
the ferromagnetic CrI$_{3}$ and CrGeTe$_{3}$ monolayers, the leading
contribution to the spin-anisotropic exchanges is from the SOC on
the nonmagnetic $5p$ Iodine or Tellurium anions rather than from
the magnetic $3d$ Cr ions because the atomic number $Z_{I}=53$ and
$Z_{Te}=52$ are much larger than that of Cr ions with $Z_{cr}=24$.
Although the nearest neighbor DM interaction vanishes due to the inversion
lattice symmetry the DM term may appear between the next-nearest neighbors \cite{Chen2018}.
It is also intriguing to revisit the magnetic properties in iron-based
superconductors since the atomic number of the surrounding nonmagnetic
anions is comparable to that of the irons \cite{Hosono2008}.

Finally, although we neglect the direct hoppings between magnetic
ions, it is straightforward to include them in our formula \cite{Chen2008}.

To conclude, we have presented an analytical study of the superexchange
of magnetic cations through nonmagnetic anions with SOC. We show that
the SOC on nonmagnetic ligands could induce the anisotropic exchange,
DM interaction and single-ion anisotropy on their neighboring magnetic
ions. The nonmagnetic ligands contribute to magnetic anisotropy in
a similar way as the magnetic ions. Our work demonstrates that exotic
quantum states in condensed matter systems, order phases in low dimensional
systems and single-molecular magnetic device could be engineered by
the SOC on nonmagnetic ligands.

\textit{Acknowledgements} We are thankful to Yuehua Su and Myung Joon Han for
fruitful discussions. This work is supported by the National Natural Science Foundation
of China (91750111, 11874188, U1930401 and 11874075), and the Fundamental Research
Funds for the Central Universities, China (No. GK201801009, GK201402011, ), National Key Research and Development Program of China (2018YFA0305703) and Science Challenge Project (TZ2016001).

\end{document}